\documentclass[aps,prb,showpacs,twocolumn]{revtex4}
\usepackage{epsfig,graphicx}
\usepackage{amsmath}
\usepackage{amsfonts}
\usepackage{amssymb}
\begin{document}

\title{Hybrid functionals for solids with an optimized Hartree-Fock mixing parameter}

\author{David Koller}
\author{Peter Blaha}
\affiliation{Institute of Materials Chemistry, Vienna University of Technology,
Getreidemarkt 9/165-TC, A-1060 Vienna, Austria}
\author{Fabien Tran}
\affiliation{Institute of Physical Chemistry, University of Zurich,
Winterthurerstrasse 190, CH-8057 Zurich, Switzerland}

\begin{abstract}

(Screened) hybrid functionals are being used more and more for solid-state
calculations. Usually the fraction $\alpha$ of Hartree-Fock exchange is kept
fixed during the calculation, however there is no single (universal) value for
$\alpha$ which systematically leads to satisfying accuracy. Instead, one could
use a 
property of the system under consideration to determine
$\alpha$ and in this way the functional would be more flexible and potentially
more accurate. Recently, it was proposed to use the static
dielectric constant $\varepsilon$ for the calculation of $\alpha$
[Shimazaki and Asai, Chem. Phys. Lett. \textbf{466}, 91 (2008) and
Marques \textit{et al}., Phys. Rev. B \textbf{83}, 035119 (2011)].
We explore this idea further and propose a scheme where the connection between 
$\varepsilon$ and $\alpha$ is optimized based on experimental
band gaps. $\varepsilon$, and thus $\alpha$, is recalculated at each iteration of
the self-consistent procedure. We present results for the band gap and lattice
constant of various semiconductors and insulators with this procedure. 
In addition, we show that this approach can also be combined with a
non-self-consistent hybrid approximation to speed up the calculations
considerably, while retaining an excellent accuracy in most cases.

\end{abstract}


\maketitle

\section{\label{Introduction}Introduction}

Hybrid functionals,\cite{becke-hybrid-1,becke-hybrid-2}
and in particular the screened versions,\cite{exp-screen,HSE,tran-hybrid} provide, in
combination with a generalization (Ref. \onlinecite{gks}) of the
Kohn-Sham equations,\cite{ks} an efficient way to deal with the well-known band
gap problem\cite{perdew-gap,orb-dep-func} in density functional theory (DFT).
\cite{hohkohn} In hybrid functionals, (semi)local
[i.e., local density (LDA) or generalized gradient approximation (GGA)] and
Hartree-Fock (HF) exchange are
mixed, and since the trends of pure semilocal and pure HF methods is to underestimate
and overestimate band gaps, respectively, the mixing of both usually leads
to more accurate band gaps. The use of hybrid functionals has been justified by
(more or less) formal arguments
(Refs. \onlinecite{becke-hybrid-1} and \onlinecite{PBE0}).

One well-established screened hybrid functional for solids is the one proposed
by Heyd, Scuseria, and Ernzerhof\cite{HSE} (HSE), which is based on the GGA
functional of Perdew, Burke, and Ernzerhof\cite{PBE} (PBE):
\begin{equation}
E_{\text{xc}}=
\alpha E_{\text{x}}^{\text{SR,HF}} +
(1-\alpha)E_{\text{x}}^{\text{SR,PBE}} +
E_{\text{x}}^{\text{LR,PBE}} +
E_{\text{c}}^{\text{PBE}},
\label{eqHSE}
\end{equation}
where $\alpha$ ($\in [0, 1]$) is the mixing parameter for the short-range exchange
(in screened hybrid functionals the long-range exchange is pure semilocal). The
indices ``x'' and ``c'' denote the exchange and correlation energy
contributions.
The HSE functional is usually used with $\alpha=0.25$.
Another parameter in Eq. (\ref{eqHSE}) is the screening parameter $\lambda$
in the error function ($\lambda=0.11$ bohr$^{-1}$ in the version
HSE06\cite{KrukauJCP06}),
which determines the separation between short-range
and long-range exchange.
A larger amount $\alpha$ of HF exchange increases the band gap, while a
larger value for $\lambda$ (i.e., more screening) decreases it.
Actually, this means that depending on the choice of these two
parameters more or less any desired result between PBE and HF can be obtained.
Furthermore, it also means that a reasonable way to fix the parameters is
very important.
In the HSE06 functional, the first of these parameters ($\alpha$) was chosen on
the basis of theoretical considerations,\cite{PBE0} whereas the second one
($\lambda$) was fitted to experimental results.\cite{HSE}
It has been shown that by fitting both
parameters further improvement can be achieved.\cite{HSEparameter}

At this point, it is worth recalling that already in 1990 Bylander and Kleinman
\cite{exp-screen} proposed an LDA-based functional containing screened HF
exchange (sX-LDA):
\begin{equation}
E_{\text{xc}}=
E_{\text{x}}^{\text{SR,HF}} +
E_{\text{x}}^{\text{LR,LDA}} +
E_{\text{c}}^{\text{LDA}},
\label{eqsXLDA}
\end{equation}
which can be regarded as a hybrid functional with $\alpha=1$. In the sX-LDA
functional, the long-range and short-range exchange are split using
the exponential function (Yukawa potential).

Up to now, the majority of calculations with hybrid
functionals (full-range, short-range, or long-range)
have been done with fixed
values for $\alpha$ and $\lambda$.
However, allowing $\alpha$ or $\lambda$ to depend on a property of the
system is a way to make the functional more flexible and thus potentially
more accurate. A brief summary of such schemes is mentioned below.
The sX-LDA functional is usually used with a screening parameter $\lambda$
which is calculated using the average of the valence electron density
(see, e.g., Ref. \onlinecite{ClarkPRB10} for recent calculations).
In Ref. \onlinecite{JaramilloJCP03}, it was proposed to make
$\alpha$ position-dependent by using the electron density $\rho$,
its derivative $\nabla\rho$, and
the kinetic-energy density, while in
Ref. \onlinecite{KrukauJCP08}, the use of a
position-dependent $\lambda$ which depends on $\rho$ and $\nabla\rho$
has been proposed.
Shimazaki and Asai \cite{ShimazakiCPL08,ShimazakiJCP09,ShimazakiJCP10}
proposed several functionals in
which either both $\alpha$ and $\lambda$ or only $\alpha$ are determined
using the static dielectric constant $\varepsilon_{s}$.
In the method presented in
Ref. \onlinecite{SteinPRL10} by Stein \textit{et al}., the screening parameter is tuned
such that the Koopmans' theorem
(which requires the use of orbital eigenvalues and total energies)
is obeyed as closely as possible.
Marques \textit{et al}. \cite{hybrid-grr} considered
two ways for the calculation of $\alpha$: either
with the static dielectric constant $\varepsilon_{s}$ or with the average of
$\left\vert\nabla\rho\right\vert/\rho$ in the unit cell\cite{hybrid-grr}
(as done originally in a similar context for the modified Becke-Johnson
exchange potential\cite{mBJ}).
We also mention that interesting discussions about the link between
screened hybrid functionals and quasiparticle theories can be found in
Refs. \onlinecite{hybrid-grr} and \onlinecite{HSEparameter}.

In the present work, we further explore the use of the dielectric
constant for the calculation of the fraction of HF exchange in
the screened hybrid functional YS-PBE0,\cite{tran-hybrid} which is based on
the Yukawa operator. We will show the results obtained with this scheme
for the band gap and lattice constant of solids.
Our work is organized as follows. In Sec. \ref{Details} the computational
details are given, while the description of our method and the results are
presented in Sec. \ref{Results}. Finally, the summary of our work will be given in
Sec. \ref{Conclusion}.

\section{\label{Details}Computational Details}

All calculations were performed with the WIEN2k software,\cite{WIEN2k} which is
based on the full potential (linearized) augmented plane wave and local
orbitals method\cite{Singh} for quantum calculations on periodic
systems. Recently, unscreened and screened hybrid functionals were implemented
into WIEN2k.\cite{tran-hybrid}
In screened hybrid functionals, the screening of the Coulomb operator is done by
using the exponential function, and it was shown (Ref. \onlinecite{tran-hybrid})
that by choosing carefully
the screening parameter $\lambda$, the results are very close to the
results from the error function-based screened hybrid functionals (e.g., HSE06).
More specifically, the screening parameter used in the exponential function
should be about $1.5$ times larger than the one used for the error function
(see Refs. \onlinecite{ShimazakiCPL08,ShimazakiJCP09,ShimazakiJCP10,tran-hybrid}
for details).
The results presented in the present work were obtained
with the YS-PBE0 functional which is based on the PBE functional\cite{PBE}
and was used in our previous works.\cite{BotanaPRB12,TranPRB12}
The YS-PBE0 calculations were done with a fixed screening parameter of
$\lambda=0.165$ bohr$^{-1}$, which gives results close to the results
obtained with the HSE06 functional.\cite{KrukauJCP06}
Spin-orbit coupling has not been considered explicitly in hybrid functional
calculations, but its influence on the band gap is considered to be the same 
as for the PBE functional and the gaps have been corrected accordingly.

The imaginary part of the dielectric function $\varepsilon$ was calculated using Fermi's
golden rule and the independent particle approximation. The Kramers-Kronig transformation was used to obtain the real
part of $\varepsilon$. Details of this approach can
be found in Ref. \onlinecite{ambrosch-draxl-optik}. This gives the $G=G'=0$
element of the dielectric matrix from which the static macroscopic dielectric
constant (which is a tensor since solids are anisotropic) can be obtained by
taking the limit $\omega\rightarrow0$ and $\mathbf{q}\rightarrow0$ of the real
part. For solids whose symmetry leads to vanishing off-diagonal elements of this
tensor we take the geometric mean of the diagonal elements of this tensor and
from now on we will refer to it simply as the dielectric constant and use the
symbol $\varepsilon^*$. Systems with non-vanishing off-diagonal elements were
not considered in the present work. All parameters of the calculations, such as
basis-set size or Brillouin zone sampling were tested for convergence. Except
for the calculations of the lattice constants, the experimental structures have
been used.

\section{\label{Results}Results and Discussion}

\subsection{\label{ResultsA}Band gap}

\begin{table*}
\caption{\label{Proben}
Fundamental band gaps (in eV) of 24 solids categorized into four different
groups: ionic compounds, $sp$-semiconductors, transition-metal oxides (labeled
as ``TmO''), and other transition-metal compounds (labeled as ``TmX'').
The Strukturbericht symbols are indicated in parenthesis. The three columns
``YS-PBE0'' show the results obtained with the hybrid functional YS-PBE0 with
$\alpha=0.25$ (Sec. \ref{ResultsA}), $\alpha_{\text{opt}}$ (Sec. \ref{ResultsA}),
and $\alpha_{\text{opt}}$ with the non self-consistent diagonal-HF approximation (Sec. \ref{ResultsC}).
The experimental band gaps (see Ref. \onlinecite{mBJ-improve} for the
references) are shown for comparison. In addition,
$\alpha_{\text{opt}}$ (obtained from the fitting procedure), and the static dielectric functions obtained from YS-PBE0
($\alpha_{\text{opt}}$), PBE, and experiment are also shown.}
\begin{ruledtabular}
\begin{tabular}{lccccccccc}
Solid & Type & YS-PBE0 ($\alpha=0.25$) & YS-PBE0 ($\alpha_{\text{opt}}$) &
YS-PBE0-diag ($\alpha_{\text{opt}}$) & exp. & $\alpha_{\text{opt}}$ &
$\varepsilon^*_{\alpha_{\text{opt}}}$ & $\varepsilon^*_{\text{PBE}}$ & $\varepsilon_{\text{exp}}$\\
\hline
LiF (B1) & ionic & 11.4 & 14.5 & 15.1 & 14.2 & 0.58 & 1.5 & 2.1 & 1.9\footnotemark[1] \\
NaF (B1) & ionic & 8.4 & 11.9 & 12.5 & 11.7 & 0.63 & 1.3 & 1.8 & 1.7\footnotemark[2] \\
KF (B1) & ionic & 8.1 & 11.0 & 11.8 & 10.9 & 0.59 & 1.4 & 2.0 & 1.8\footnotemark[2] \\
LiCl (B1) & ionic & 7.8 & 9.0 & 8.9 & 9.4 & 0.45 & 2.1 & 3.3 & 2.7\footnotemark[2] \\
NaCl (B1) & ionic & 6.5 & 7.9 & 7.9 & 8.6 & 0.50 & 1.8 & 2.7 & 2.3\footnotemark[3] \\
KCl (B1) & ionic & 6.5 & 8.0 & 8.1 & 8.5 & 0.52 & 1.7 & 2.6 & 2.2\footnotemark[2] \\
CaO (B1) & ionic & 5.3 & 6.5 & 7.2 & 7.0\footnotemark[10] &
0.43 & 2.3 & 3.9 & 3.3\footnotemark[11] \\
C (A4) & $sp$ & 5.4 & 5.5 & 5.5 & 5.48 & 0.30 & 4.0 & 5.8 & 5.7\footnotemark[1] \\
Si (A4) & $sp$ & 1.16 & 1.08 & 1.07 & 1.17 & 0.22 & 8.2 & 12.9 & 11.9\footnotemark[1] \\
Ge (A4) & $sp$ & 0.77 & 0.61 & 0.53 & 0.74 & 0.20 & 11.4 & 20.8 & 15.9\footnotemark[3] \\
CdS (B3) & $sp$ & 2.12 & 2.42 & 2.44 & 2.42 & 0.32 & 3.7 & 6.3 & 5.3\footnotemark[1] \\
GaN (B3) & $sp$ & 2.8 & 3.2 & 3.2 & 3.2 & 0.32 & 3.6 & 6.0 & 5.3\footnotemark[1] \\
BN (B3) & $sp$ & 5.8 & 6.4 & 6.3 & 6.25 & 0.35 & 3.2 & 4.6 & 4.5\footnotemark[1] \\
SiC (B3) & $sp$ & 2.25 & 2.40 & 2.38 & 2.4 & 0.29 & 4.6 & 7.0 & 6.5\footnotemark[1] \\
AlP (B3) & $sp$ & 2.29 & 2.35 & 2.33 & 2.45 & 0.27 & 5.4 & 8.5 & 7.5\footnotemark[1] \\
InP (B3) & $sp$ & 1.42 & 1.39 & 1.37 & 1.42 & 0.24 & 6.7 & 10.9 & 9.6\footnotemark[3] \\
Cu$_2$O (C3) & TmO & 1.89 & 2.25 & 2.51 & 2.17 & 0.31 & 3.8 & 9.3 & 6.5\footnotemark[4] \\
TiO$_2$ (C4) & TmO & 3.3 & 3.7 & 5.5 & 3.3 & 0.31 & 4.0 & 7.9 & 6.3\footnotemark[5] \\
ZnO (B4) & TmO & 2.5 & 3.7 & 4.5 & 3.44 & 0.42 & 2.4 & 4.9 & 3.7\footnotemark[1] \\
SrTiO$_3$ (E2$_1$) & TmO & 3.2 & 3.8 & 5.6 & 3.25 & 0.34 & 3.4 & 6.3 & 5.2\footnotemark[6] \\
MnO (B1) & TmO & 2.9 & 3.8 & 4.6 & 3.9 & 0.37 & 2.8 & 7.7 & 5.0\footnotemark[7] \\
ScN (B1) & TmX & 0.84 & 0.84 & ---\footnotemark[9] & 0.9 & 0.25 & 5.9 & 12.1 & 7.2\footnotemark[8] \\
MoS$_2$ (C7) & TmX & 1.44 & 1.33 & 1.32 & 1.29 & 0.22 & 9.0 & 13.9 & \\
ZnS (B3) & TmX & 3.3 & 3.7 & 3.7 & 3.91 & 0.32 & 3.7 & 6.2 & 5.1\footnotemark[1] \\
\end{tabular}
\end{ruledtabular}
\footnotetext[1]{Reference \onlinecite{epsilon-LiF}.}
\footnotetext[2]{Reference \onlinecite{epsilon-NaF}.}
\footnotetext[3]{Reference \onlinecite{epsilon-NaCl}.}
\footnotetext[4]{Reference \onlinecite{epsilon-Cu2O}.}
\footnotetext[5]{Reference \onlinecite{epsilon-TiO2}.}
\footnotetext[6]{Reference \onlinecite{epsilon-SrTiO3}.}
\footnotetext[7]{Reference \onlinecite{epsilon-MnO}.}
\footnotetext[8]{Reference \onlinecite{epsilon-ScN}.}
\footnotetext[9]{It is not recommended to apply the diagonal-HF approximation to ScN
since it is metallic with PBE.}
\footnotetext[10]{Reference \onlinecite{gap-CaO}.}
\footnotetext[11]{Reference \onlinecite{Madelung-semiconductors}.}
\end{table*}

\begin{figure}[htb]
\centering \includegraphics[angle=0,width=\columnwidth]{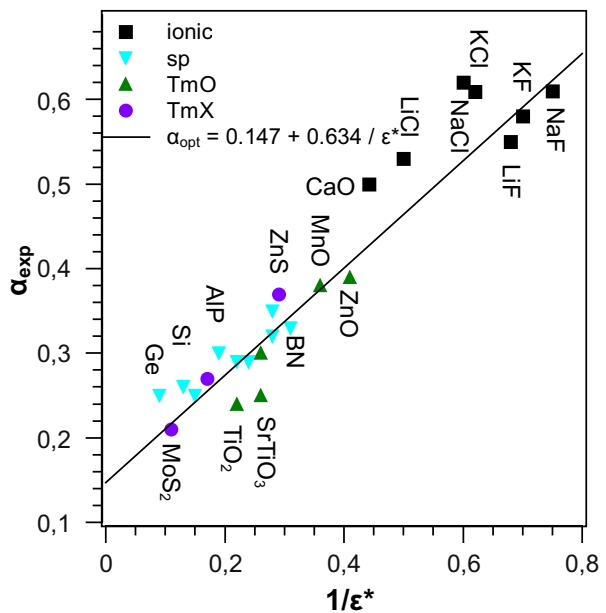}
\caption{\label{Fig1}(Color online) The fraction
of HF exchange required to reproduce the experimental band gap
($\alpha_{\text{exp}}$) versus
the inverse of the dielectric constant
($1/\varepsilon^*$) which is obtained from a YS-PBE0 calculation using the
corresponding $\alpha_{\text{exp}}$.
The straight line [$\alpha_{\text{opt}}=\alpha_{\text{opt}}(\varepsilon^*)$] is
a fit corresponding to the least square error of the gap and not of $\alpha$.}
\end{figure}

\begin{figure}[htb]
\centering \includegraphics[angle=0,width=\columnwidth]{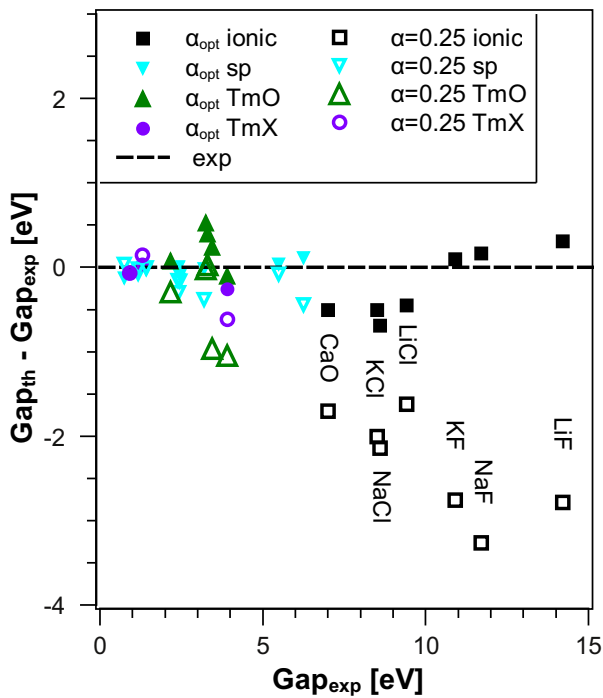}
\caption{\label{Fig2}(Color online) Difference between calculated (full
symbols: YS-PBE0 with $\alpha_{\text{opt}}$, open symbols:
YS-PBE0 with $\alpha=0.25$) and experimental band gaps.
The shape of the symbol indicates the type of solids.}
\end{figure}

\begin{figure}[htb]
\centering \includegraphics[angle=0,width=\columnwidth]{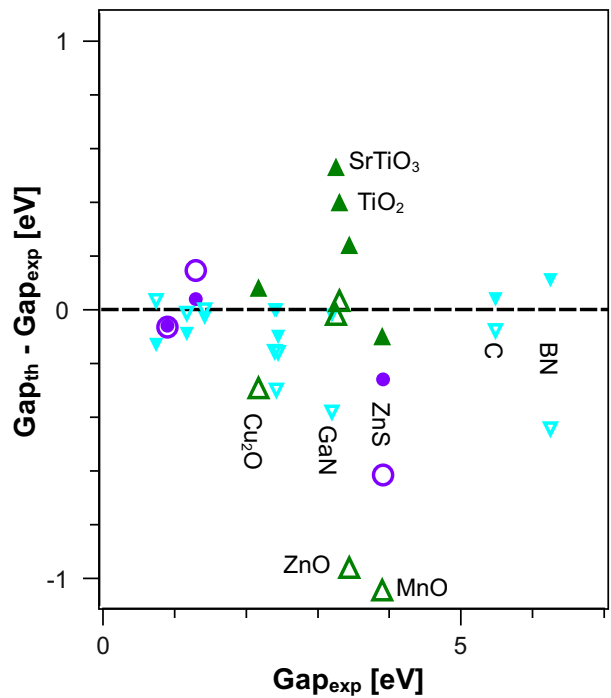}
\caption{\label{Fig3}(Color online) Same as Fig. \ref{Fig2}, but
only for solids with band gaps smaller than 7 eV.}
\end{figure}

\begin{figure}[htb]
\centering \includegraphics[angle=0,width=\columnwidth]{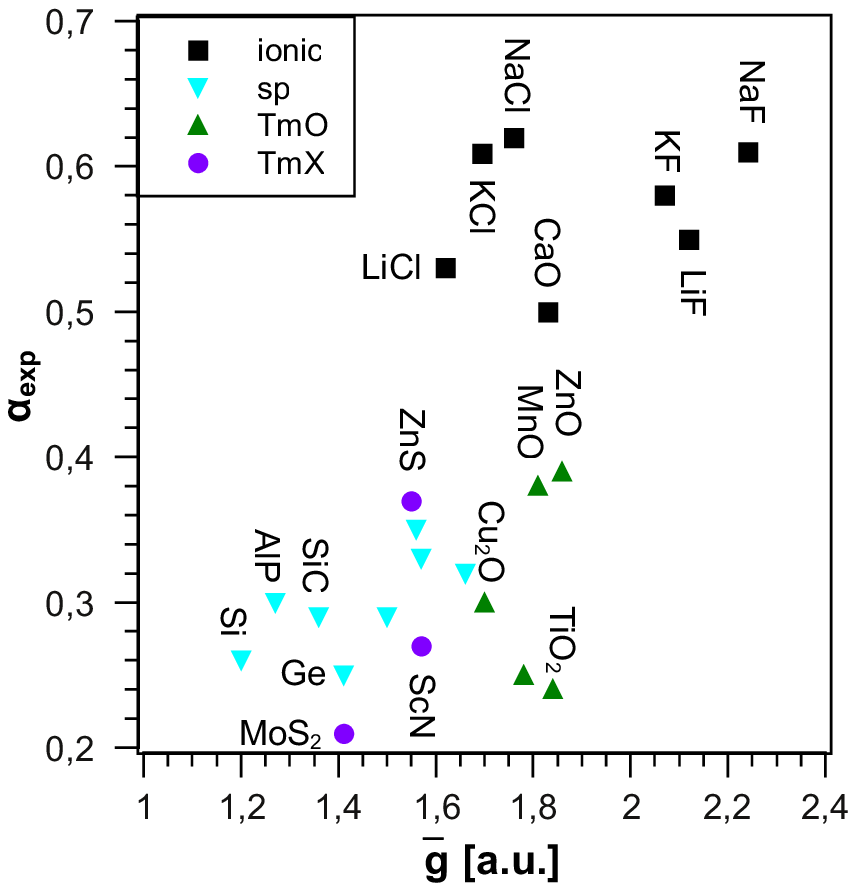}
\caption{\label{Fig4}(Color online) The fraction
of HF exchange required to reproduce the experimental band gap
($\alpha_{\text{exp}}$) versus $\overline{g}$ (see text for
definition).}
\end{figure}

Hybrid functionals with fixed amount of exact exchange and screening length can
be problematic if applied to different types of systems. For instance HSE is
known\cite{hybrid-overview} to work well for many small-gap semiconductors but
strongly underestimates the gaps of highly ionic compounds such as NaCl.
This means that in order to find an hybrid functional with broader
applicability, it is important to include different types of solids in the
fitting set. The solids that we considered for this study are listed in Table
\ref{Proben}. They include highly ionic compounds with large band gaps,
technologically important $sp$-semiconductors and more or less strongly
correlated transition-metal compounds which can be split further into
oxides (TmO) and non-oxides (TmX).

As already mentioned, the dielectric constant $\varepsilon^*$ will be used
for the determination of the fraction $\alpha$ of HF exchange. Note that in the
present work, the screening parameter $\lambda$
has been kept fixed at $\lambda=0.165$ bohr$^{-1}$
(Ref. \onlinecite{tran-hybrid}) to make comparison with YS-PBE0 (and therefore
HSE06) possible. As we can see
from Fig. \ref{Fig1}, choosing $\varepsilon^*$ seems to be a very good choice.
Indeed, there is a nice linear correlation
between $\alpha_{\text{exp}}$, which is the amount of HF exchange required to
reproduce the experimental band gap and $1/\varepsilon^*$ which is
obtained from a YS-PBE0 calculation using the corresponding $\alpha_{\text{exp}}$.
We considered the linear relation $\alpha_{\text{opt}}=A+B/\varepsilon^*$ and a
fit procedure minimizing the least-square error in the gaps leads to $A=0.147$
and $B=0.634$.
Note, that our linear approximation does not go through the origin and the
slope is not one, while in a previous work  reported in Ref. \onlinecite{hybrid-grr} a strict proportionality
$\alpha = 1 / \varepsilon^{\text{PBE}}$ was used (in unscreened PBE0).
The linear fit is shown in Fig. \ref{Fig1}, where we can see that most data
points are quite close to it. The exceptions are the ionic chlorides and the
noble gases (not shown explicitly) whose
$\alpha_{\text{exp}}$ are larger than the values $\alpha_{\text{opt}}$ obtained
from the fit and rutile and strontium titanate, where $\alpha_{\text{exp}}$ is
a bit smaller than the fitted $\alpha_{\text{opt}}$. Thus,
it can be expected that the gaps of the chlorides will be underestimated, 
whereas the gaps of rutile and strontium titanate will be overestimated.

The resulting band gaps are listed in Table \ref{Proben} and are compared with
the values obtained with the standard fixed value $\alpha=0.25$. The band gaps
are also plotted in Figs. \ref{Fig2} and \ref{Fig3} for a better comparison.
It can be clearly seen that in nearly all cases, improved band gaps are
obtained with $\alpha_{\text{opt}}$ compared to
$\alpha=0.25$, and in some cases the improvement is quite impressive.
In particular, the band gaps of the highly ionic
compounds, which are known\cite{hybrid-overview} to be underestimated by about
3 eV in HSE, are
reproduced much more accurately with the $\varepsilon^*$-dependent
$\alpha_{\text{opt}}$. As expected from
Fig. \ref{Fig1}, the use of $\alpha_{\text{opt}}$ leads to underestimations
(which are much smaller than when HSE is used) for
the chlorides and overestimations in the case of rutile and strontium
titanate. For the other solids the performance can be considered as
excellent. The deviation from experiment is within a range of $\pm0.3$ eV which
is quite good, since experimental errors and temperature effects need to be considered as well. The
performance is even slightly better than that of the modified Becke-Johnson potential
with the parameters suggested in Ref. \onlinecite{mBJ-improve}.

Since the static dielectric constant $\varepsilon^*$ is a central quantity in the described
approach, it is important to compare the calculated values to the experimental results. Table
\ref{Proben} shows the values obtained from calculations (with PBE and
YS-PBE0$_{\alpha_{\text{opt}}}$) and experiment.
We can see that YS-PBE0 with
$\alpha_{\text{opt}}$ underestimates the experimental values by about 30\%,
contrary to PBE which slightly overestimates them. This is not surprising, since gaps
are underestimated by PBE and lower gaps lead to higher dielectric
constants. In our calculations, however, the dielectric function is calculated
in the independent particle approximation and does not include
electron-hole interactions, which would lead to an increase in the dielectric
constants.\cite{Kresse-vertex} An improved calculation for  $\varepsilon^*$
would require a re-parameterization of $\alpha_{\text{opt}}$.

As a last remark in this section, we also mention that
an alternative way to determine $\alpha$ would be to use
\begin{equation}
\overline{g} = \frac{1}{V_{\text{cell}}}\int\limits_{\text{cell}}
\frac{\left\vert\nabla\rho(\mathbf{r})\right\vert}
{\rho(\mathbf{r})}d^{3}r,
\label{g-average}
\end{equation}
which is the average of
$\left\vert\nabla\rho\right\vert/\rho$ in the unit cell and has been used
successfully for the modified Becke-Johnson
potential.\cite{mBJ,mBJ2,mBJ3,mBJ-improve}
However, as Fig. \ref{Fig4} shows, the situation with $\overline{g}$ is quite
different since there is no clear correlation between the values
of $\alpha_{\text{exp}}$ and $\overline{g}$. Different classes of materials
($sp$-semiconductors, ionic chlorides or fluorides, transition metal compounds)
would require drastically different parameterizations. The transition-metal
oxides show only small variations
in $\overline{g}$ but require quite different
$\alpha_{\text{opt}}$, which is of course very problematic if any kind of
relation $\alpha_{\text{opt}}=\alpha_{\text{opt}}(\overline{g})$ is desired.
This is in quite strong contrast to the findings in
Ref. \onlinecite{hybrid-grr}, where they found a good relation between $\alpha$
and $\overline{g}$. However, their set of solids was much more
restricted in terms of transition-metal or highly ionic systems, and
furthermore, their calculations were obtained with the pseudopotential plane wave
method, whereas the calculations presented in the present work were obtained from
an all-electron method, this difference leading certainely to different values of
$\overline{g}$.
Therefore, we considered only the parameterization in terms of the dielectric
constant.


\subsection{\label{ResultsB}Lattice parameters}

\begin{figure}[htb]
\centering \includegraphics[angle=0,width=\columnwidth]{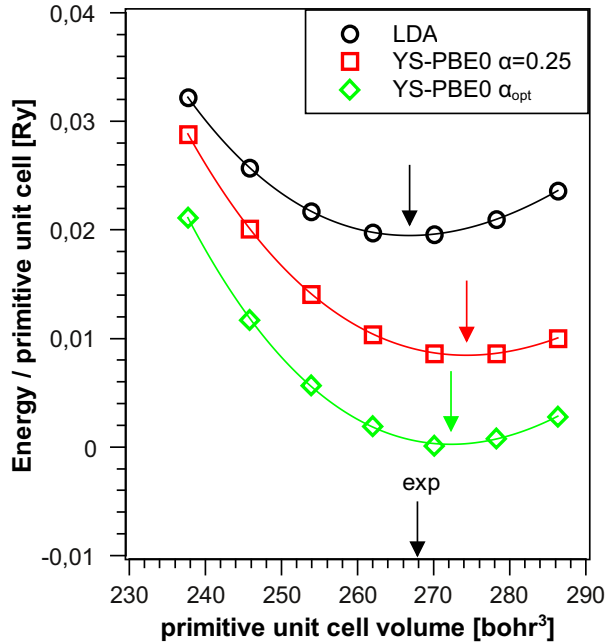}
\caption{\label{Fig5} (Color online) Total energy of silicon as a function of
  atomic volume for 3 different functionals. The lines represent fits to the
  Birch-Murnaghan-EOS, the arrows indicate the corresponding minima.}
\end{figure}

\begin{table*}
\caption{\label{struktur}
Comparison between theoretical and experimental lattice parameters (in \AA)
for several solids.
All experimental values except ScN are corrected for zero-point
anharmonic expansion.\cite{Haas-lattice} The values of $\alpha_{\text{opt}}$
and $d\alpha_{\text{opt}}/d\ln V$ at the experimental volume are also shown.}
\begin{ruledtabular}
\begin{tabular}{lcccccccc}
Solid & exp.                & LDA                  & WC                   &
HSE06                & YS-PBE0 & YS-PBE0 & $\alpha_{\text{opt}}(V=V_{\text{exp}})$ & $\frac{d\alpha_{\text{opt}}}{d\ln V}(V=V_{\text{exp}})$ \\
 &                 &                   & &
                & ($\alpha=0.25$) & ($\alpha_{\text{opt}}$) & & \\

\hline
C  & 3.54\footnotemark[1] & 3.54\footnotemark[1] & 3.56\footnotemark[1] & 3.55\footnotemark[2] & 3.55 & 3.55 & 0.306 & $-0.025$ \\
Si    & 5.42\footnotemark[1] & 5.41\footnotemark[1] & 5.44\footnotemark[1] & 5.44\footnotemark[2] & 5.46 & 5.44 & 0.224 & $-0.015$ \\
SiC  & 4.34\footnotemark[1] & 4.33\footnotemark[1] & 4.36\footnotemark[1] & 4.35\footnotemark[2] & 4.36 & 4.34 & 0.284 & $-0.041$ \\
BN  & 3.59\footnotemark[1] & 3.59\footnotemark[1] & 3.61\footnotemark[1] & 3.60\footnotemark[2] & 3.61 & 3.60 & 0.346 & $-0.021$ \\
GaN  & 4.52\footnotemark[1] & 4.46\footnotemark[1] & 4.50\footnotemark[1] & 4.49\footnotemark[2] & 4.52 & 4.50 & 0.323 & $-0.098$ \\
ScN   & 4.50\footnotemark[3] & 4.43\footnotemark[3] & 4.47\footnotemark[3] &                      & 4.51 & 4.52 & 0.254 & 0.042 \\
CaO  & 4.79\footnotemark[1] & 4.72\footnotemark[1] & 4.78\footnotemark[1] &                      & 4.82 & 4.86 & 0.427 & 0.077 \\
LiCl  & 5.07\footnotemark[1] & 4.97\footnotemark[1] & 5.07\footnotemark[1]&5.12\footnotemark[2] & 5.15 & 5.55 & 0.447 & 0.15 \\
NaCl  & 5.60\footnotemark[1] & 5.48\footnotemark[1] & 5.62\footnotemark[1] & 5.64\footnotemark[2] & 5.69 & 6.30 & 0.498 & 0.13 \\
\end{tabular}
\end{ruledtabular}
\footnotetext[1]{Reference \onlinecite{Haas-lattice}.}
\footnotetext[2]{Reference \onlinecite{Kresse-lattice}.}
\footnotetext[3]{Reference \onlinecite{Tran-WC}.}
\end{table*}


Another important test for functionals is the total energy, in particular
how  accurate equilibrium structural parameters  or atomization
energies can be described. Note, that $\alpha_{\text{opt}}$ will change as function of
the lattice parameter and this could have important consequences on the
equilibrium geometry. Therefore, we calculated the equilibrium lattice
parameters  for a few selected solids by calculating the total energies
for different lattice parameters and fit them to  
the Birch-Murnaghan\cite{birch-murnaghan} equation of state (EOS). The minimum of this fit
corresponds to the equilibrium lattice parameter. An example is shown in
Fig. \ref{Fig5} for silicon and smooth total energy fits can be obtained for all
functionals, including YS-PBE0 with $\alpha_{\text{opt}}$. 

Table \ref{struktur} shows the equilibrium lattice
parameters obtained with the hybrid functional
YS-PBE0 for a few representative solids. For comparison, the
results obtained with other functionals, namely, LDA, Wu-Cohen
(WC),\cite{WuPRB06} and HSE06, as well as from experiment are also shown.
The selected solids include cases with low (e.g., LiCl), intermediate
(e.g., ScN), and high (e.g., Si) values of the dielectric constant.
As already known,\cite{Tran-WC,Haas-lattice,Kresse-lattice}
LDA underestimates strongly the lattice constants,
while WC belongs to the group of the most accurate GGA functionals for this
property, and HSE06 has the tendency to give too large values (albeit not as
much as PBE). In most cases YS-PBE0 with fixed $\alpha=0.25$ is in excellent
agreement with HSE06, while for NaCl a slightly larger deviation is observed.
The results for YS-PBE0 with $\alpha_{\text{opt}}=A+B/\varepsilon^*$ are in
most cases very similar to those with fixed $\alpha$ and thus one could
conclude that a hybrid functional with optimized $\alpha$ is well suited also
for lattice parameter determinations. However, we can see that for the
highly ionic compounds LiCl and NaCl, a dramatic increase of
the equilibrium lattice parameters, reaching completely unphysical values (an
increase from 5.15 to 5.55 \AA~ and 5.69 to 6.20  \AA~ for LiCl and NaCl,
respectively), is obtained. Clearly this
approach is not recommendable in such a case.
In order to know where the problem comes from it is necessary to look more
closely at the volume dependence of $\alpha_{\text{opt}}$ (see Table \ref{struktur}).
In the case of CaO and especially LiCl and NaCl there is a
very large slope of $\alpha_{\text{opt}}$ as function of the volume,
indicating that $\varepsilon^*$ (and the band gap) increases (decreases)
strongly with reduced volume. On the other hand, for typical semiconductors
$\alpha_{\text{opt}}$ varies much less strongly when the volume is changed and the variations
in the band gap are compensated by those in the dielectric function. Together
with the fact that $\alpha_{\text{opt}}$ is almost twice as large as
$\alpha_{\text{HSE}}=0.25$ the equilibrium lattice parameters become very large in
highly ionic materials. It should be noted, however, that it is the
variation of $\alpha_{\text{opt}}$ with volume and not the large $\alpha$ value
itself, which causes the failure to obtain good equilibrium volumes. For
instance, for LiCl a lattice parameter of 5.17 \AA~ is obtained if the
fixed value $\alpha=0.447$ (obtained at $V_{\text{exp}}$) is used.
Therefore, an approach using a fixed $\alpha$
offers an alternative way to describe simultaneously the energy band gap 
and structural parameters in a reasonable way.


Next we want to test the YS-PBE0($\alpha_{\text{opt}}$) scheme for  a  more complicated
example: the pressure-induced B1-B2 phase transition in CaO. Here the
goal is to find the transition pressure above which the
enthalpy of the B2 phase becomes lower than that of the B1
phase. Experimentally this transition is found to occur in the range of 60-70
GPa.\cite{CaO-exp} Standard DFT functionals can reproduce this quite well: LDA
predicts the transition to occur at 57 GPa, WC at 61 GPa and PBE is in the
center of the experimental range with a transition pressure of 66
GPa. Traditional YS-PBE0 with $\alpha=0.25$ also predicts it correctly at 67 GPa, but if
$\alpha_{\text{opt}}$ is used the transition pressure is
shifted to 87 GPa. This means that the result is worsened in a similar way as
the result for the CaO lattice parameter (it is however not as bad as the 0.5
\AA\ error for LiCl). The shift of the transition point to
higher pressure is related to the different $\alpha_{\text{opt}}$ values of
both phases (B1: 0.43, B2: 0.39). The higher $\alpha_{\text{opt}}$ of the
B1-phase reduces its total energy so that more pressure is required for a
transition to the B2 phase.

\subsection{\label{ResultsC} Combining the optimized-$\alpha$-approach with
the diagonal-hybrid-approximation}

Since hybrid calculations are
computationally demanding, it is desirable to have a way to get similar
results to the approach described above in a much shorter time. This is made possible
by the non self-consistent diagonal-only hybrid approximation proposed in
Ref. \onlinecite{tran-diagonal}. In this approximation, only the
diagonal elements of the perturbation Hamiltonian (hybrid minus semilocal)
are calculated (in the basis of the semilocal orbitals), while the non-diagonal 
elements are neglected. 
This saves the time of evaluating most of the matrix elements and also a
self-consistency cycle is not necessary. It was found
previously\cite{tran-diagonal} that for common semiconductors and insulators
this procedure leads to gaps in very close agreement with fully self-consistent
calculations. 
In addition we assume a linear dependency of the gap and  $1/\varepsilon^*$ on
$\alpha$, so that we can obtain $\alpha_{\text{opt}}$ and the corresponding
band gap from two calculations at  $\alpha_1=0$ (plain PBE) and $\alpha_2=0.3$.

The results obtained by this procedure are included in
Table \ref{Proben}. Except for the transition metal oxides (and to a much
lesser degree the ionic fluorides),
the deviation from the full-hybrid gaps is quite small, usually in the range of
a few hundredth of an eV. On the other hand, the reduction of the computational
effort is substantial, which allows to apply this approximation also to large
unit cells.

\begin{figure}[htb]
\centering \includegraphics[angle=0,width=\columnwidth]{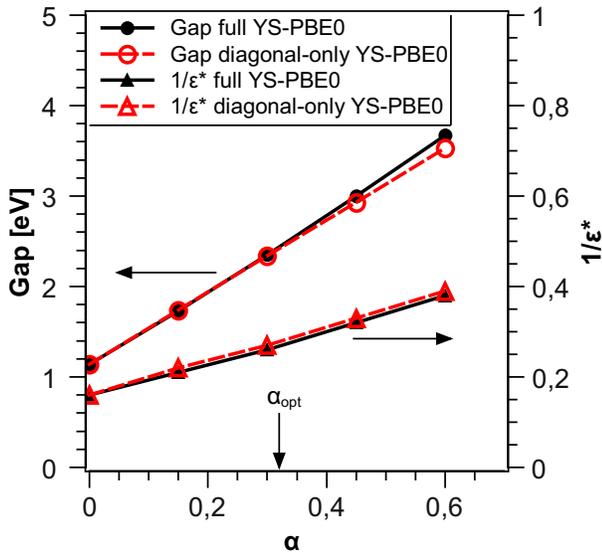}
\caption{\label{Fig6} (Color online) Dependence of the band gap and the inverse
dielectric constant on the amount of HF exchange in full and diagonal
YS-PBE0 in CdS. Lines are a guide to the eye.}
\end{figure}

\begin{figure}[htb]
\centering \includegraphics[angle=0,width=\columnwidth]{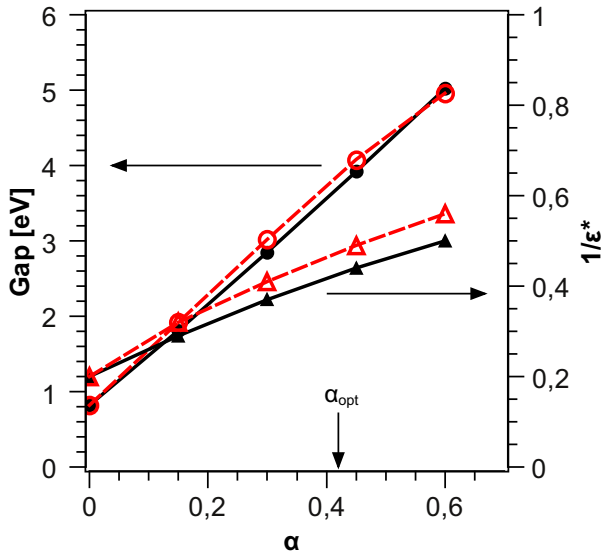}
\caption{\label{Fig7} (Color online)  Dependence of the band gap and the inverse
dielectric constant on the amount of HF exchange in full and diagonal
YS-PBE0 in  ZnO. Lines are a guide to the eye.}
\end{figure}

In order to analyze why this scheme works for most cases but has some exceptions,
we take a closer look at CdS as an example for a working case and at ZnO as an
example for an exception. The relevant data are shown in Figs. \ref{Fig6} and
\ref{Fig7}. We can see that in both cases the dependency of the band gaps
on $\alpha$ using the full
hybrid or the diagonal-hybrid approximation is nearly identical and fairly
linear.  For CdS the same holds for the dependency of $1/\varepsilon^*$  on
$\alpha$. However, for ZnO there is a substantial difference
between the dielectric constant obtained with full hybrid calculations and that
obtained with the diagonal approximation. The reason for that is that in the
diagonal calculations only the eigenvalues change (to almost the same values as
in full hybrid calculations), but the orbitals do not and still belong to the
single-particle 
Hamiltonian with a semilocal potential. While in the case of CdS the
eigenfunctions from PBE or full hybrid-DFT calculations do not  
differ much so that the dielectric constants with these two schemes remain 
very similar, in the case of ZnO
there is a strong difference in the eigenfunctions from PBE or hybrid-DFT. 
This leads to different momentum matrix elements and therefore
dielectric constants and finally the band gaps differ substantially.

\section{\label{Conclusion}Summary and Conclusion}

A hybrid functional was presented which, in contrast to most currently
established hybrid functionals, uses a HF mixing parameter which is
individually adapted for each investigated system. This is
achieved automatically using the calculated dielectric constant. Since
this is a global quantity of a bulk material, the functional
as described here can be applied reasonably only to systems with strict
3-dimensional periodicity. However it may be generalized in the course of
future work so that defect structures, molecules, surfaces and interfaces can
be dealt with. Actually, it has been claimed\cite{louie-hybrid-kritik} that a
hybrid functional with position-independent mixing factor and screening length
is not appropriate for such cases.

The results obtained with this functional are quite
impressive. Band gaps for a wide range of materials are much better reproduced than by hybrid functionals
with a fixed parameter. Structural properties, which are usually not considered in
similar studies, are in most cases of
excellent quality and only in highly ionic insulators problems appear.
However, even in these cases, good results can be obtained when one fixes $\alpha_{\text{opt}}$
at one volume and neglects its volume variations. 

We have also tested an approximate version of this approach, namely a
non self-consistent hybrid scheme considering only the diagonal HF matrix
elements. Except for transition metal oxides (where even non self-consistent $GW$
methods fail) this approach leads to almost
identical band gaps and such a method can be used to predict reliable band gaps
in systems with hundreds of atoms. 

\begin{acknowledgments}

This work was supported by the project SFB-F41 (ViCoM) of the Austrian Science Fund.

\end{acknowledgments}

\bibliography{references}

\end{document}